\documentclass[12pt,preprint]{aastex}
\usepackage{amsmath,amssymb}
\usepackage{graphicx}
\usepackage{float}
\usepackage{multirow}
\usepackage{threeparttable}
\usepackage{rotating}

\slugcomment{KSUPT-13/5 \hspace{0.5truecm} November 2013}

\begin{document}

\title{Median statistics cosmological parameter values}

\author{Sara Crandall and Bharat Ratra}

\affil{Department of Physics, Kansas State University, 116 Cardwell Hall, Manhattan, KS 66506, USA} 
                 
\email{sara1990@k-state.edu, ratra@phys.ksu.edu}

\begin{abstract}

We present median statistics central values and ranges for 12 cosmological parameters, using 582 measurements (published during 1990-2010) collected by \cite{Croft2011}. On comparing to the recent $Planck$ collaboration \citep{Ade2013} estimates of 11 of these parameters, we find good consistency in nine cases.

\end{abstract}

\maketitle

\section{Introduction}
\label{intro}
Recent cosmic microwave background anisotropy \citep[see, e.g.,][]{Hinshaw2013, Ade2013}, baryon acoustic oscillation peak length scale \citep[see, e.g.,][]{Busca2013, Farooq2013c}, supernova Type Ia apparent magnitude versus redshift \citep[see, e.g.,][]{Campbell2013, Liao2013}, and Hubble parameter as a function of redshift \citep[see, e.g.,][]{Moresco2012, Farooq2013b, Farooq2013d} measurements have small enough statistical error bars to encourage the belief that we will soon be in an era of precision cosmology. Of course, there have also been many earlier measurements, most having larger error bars, that have helped the field develop to the current position. In this paper we use statistical techniques to combine the results of the many earlier measurements, and so derive summary estimates of the corresponding cosmological parameters with much tighter error bars than any individual earlier measurement. We then compare these summary results to more precise recent measurements, largely those from the recent analysis of early $Planck$ space mission cosmic microwave background (CMB) anisotropy data \citep{Ade2013}. Using large-angle CMB anisotropy data to measure cosmological parameters is appealing because, once initial conditions and ionization history are established, it is possible to accurately compute cosmological model CMB anisotropy predictions as a function of cosmological parameter values. 

Previous CMB anisotropy experiments, such as $WMAP$ and ground-based ones, along with data from other techniques discussed above, have focussed attention on a ``standard" cosmological model \citep[for detailed discussions see][]{Hinshaw2013, Ade2013}. This model, called the $\Lambda$CDM model \citep{Peebles1984}, is a spatially-flat cosmological model with a current energy budget dominated by a time-independent dark energy density in the form of Einstein's cosmological constant $\Lambda$ that contributes $68.3\%$ of the current energy budget, non-relativistic cold dark matter (CDM) is the next largest contributor at  $26.7\%$, followed by non-relativistic baryonic matter at $4.9\%$ \citep{Ade2013}. For recent reviews see \cite{Wang2012}, \cite{Tsujikawa2013}, and \cite{Sola2013}.

A main goal of the $Planck$ mission is to measure cosmological parameters accurately enough to check consistency with the $\Lambda$CDM model, as well as to possibly detect deviations. However, it is also of interest to find out if previous estimates of cosmological parameters are consistent with the $Planck$ results. \cite{Ade2013}, and references therein, have compared the $Planck$ results to individual earlier measurements, most notably to the results from the $WMAP$ experiment, from which they find small differences. However, it is also of interest to attempt to derive summary estimates for cosmological parameters from the many earlier measurements that are available, and to compare these summary estimates to the $Planck$ results. This is what we do in this paper. 

To derive our summary estimates of cosmological parameter values we use the very impressive compilation of data of \cite{Croft2011}. We use 582 (of the 637) measurements for the dozen cosmological parameters collected by \cite{Croft2011}. These values were published during 1990-2010, and, as estimated by \cite{Croft2011}, are approximately 60\% of the measurements of the 12 cosmological parameters published during these two decades. The main focus of the \cite{Croft2011} paper was to compare earlier and more recent measurements and analyze how measuring techniques and results evolve over time.  In our paper we use two statistical techniques, namely weighted mean and median statistics, to find the best-fit summary measured value of each of the 12 cosmological parameters. We then compare our summary values to those found from the $Planck$ data. 

In the next section we briefly review the \cite{Croft2011} data compilation. Sections \ref{WA Stat} and \ref{Med Stat} are brief summaries of the weighted mean and median statistics techniques we use to analyze the \cite{Croft2011} data. Our analyses and results are described and discussed in Sec.~\ref{Analysis}, and we conclude in Sec.~\ref{Conclusion}. 

\section{Data Compilation}
\label{Croft/Dailey summary}
The data we use in our analyses here were compiled by \cite{Croft2011}. These data were collected from the abstracts of  papers listed on the NASA Astrophysics Data System (ADS)\footnote{adsabs.harvard.edu}. They estimate that by searching abstracts only, about 40$\%$ of available measurements were missed. Nevertheless, a great deal of data were collected. \cite{Croft2011} searched papers published in a 20 year period (1990-2010) and tabulated 637 measurements. Of the 637 measurements, 582 were listed with a central value and $1\sigma$ error bars (these are the data we use in this paper\footnote{Most of these measurements were listed with two significant figures, so results of our analyses are tabulated to two significant figures (except for $\omega_{0}$, which consisted mostly of three significant figure measurements and were so tabulated here). The error bar we use in our analyses is the average of the $1\sigma$ upper and lower error bars of \cite{Croft2011}.}) while 55 were upper or lower limits with no central value. 

The 12 cosmological parameters \cite{Croft2011} considered are:\\
\begin{enumerate}
\item $\Omega_{m}$, the non-relativistic matter density parameter.
\item $\Omega_{\Lambda}$, the cosmological constant density parameter.    
\item $h$, the Hubble constant in units of 100  km s$^{-1}$ Mpc$^{-1}$
\item $\sigma_{8}$, the rms amplitude of (linear) density perturbations averaged over 8 $h^{-1}$ Mpc spheres.
\item $\Omega_{b}$, the baryonic matter density parameter.
\item $n$, the primordial spectral index.
\item $\beta$ = $\Omega_{m}^{0.6}/b$, where $b$ is the galaxy bias.
\item $m_{\nu}$, the sum of neutrino masses.
\item $\Gamma$ = $\Omega_{m}h$.
\item $\Omega_{m}^{0.6}\sigma_{8}$.
\item $\Omega_{k}$, the space curvature density parameter.
\item $\omega_{0}$, the dark energy equation of state parameter in a simplified, incomplete, XCDM-like parameterization.  
\end{enumerate}

\begin{center}
\begin{figure}[H]
\advance\leftskip-1.25cm
\advance\rightskip-1.25cm
\includegraphics[width=63mm,height=58mm]{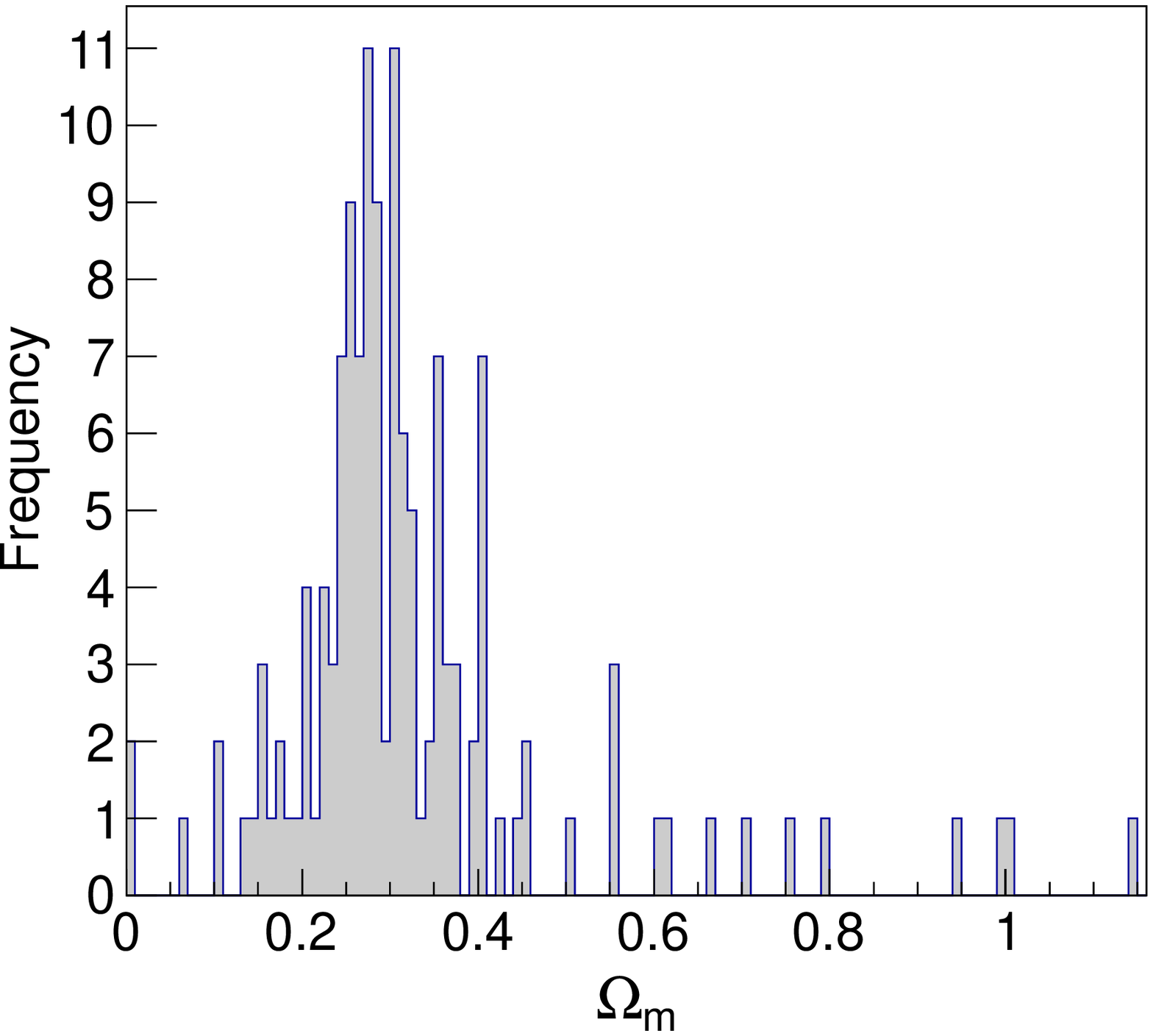}
\includegraphics[width=64mm,height=58mm]{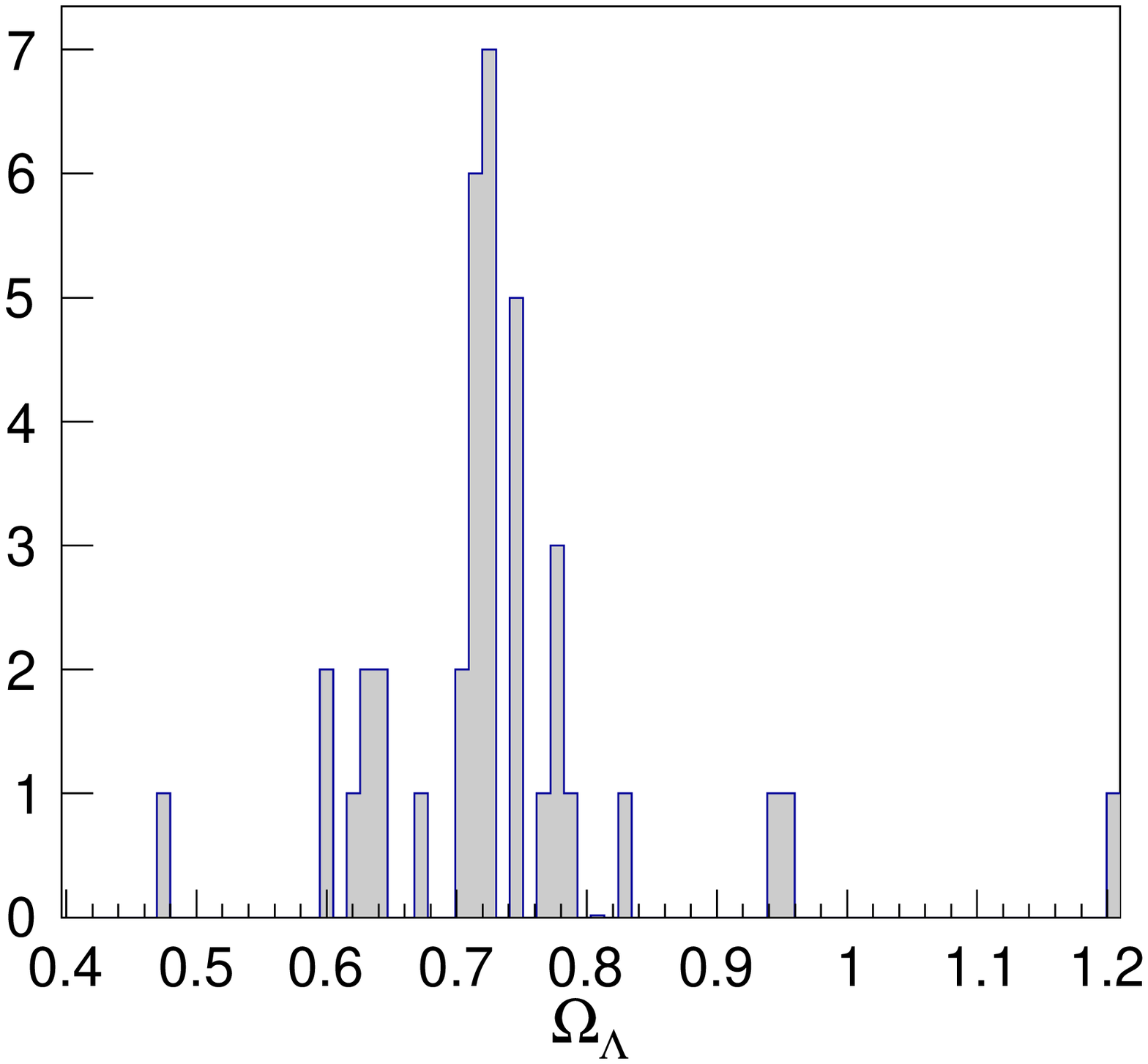}
\includegraphics[width=64mm,height=59mm]{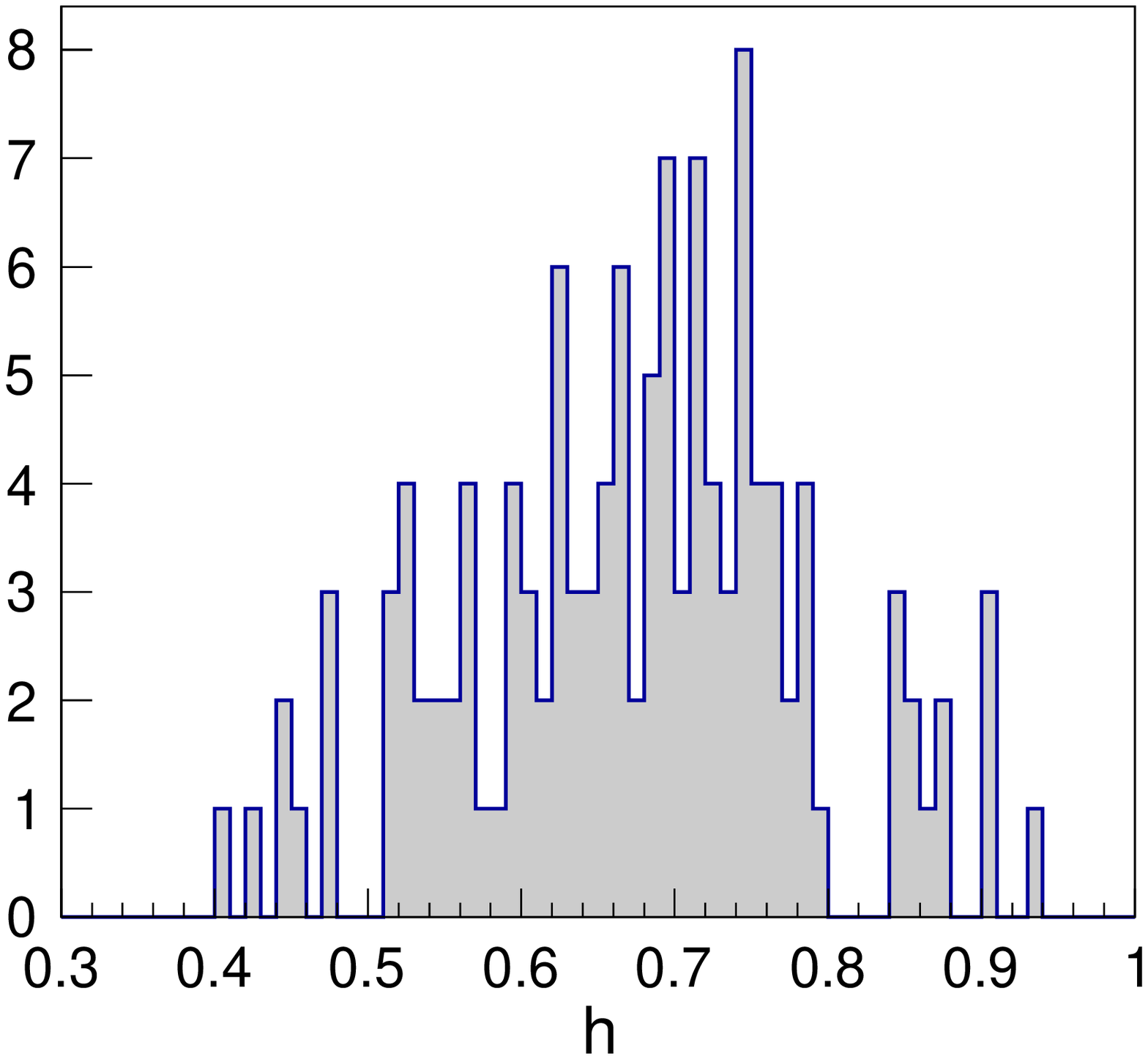}
\includegraphics[width=63mm,height=58mm]{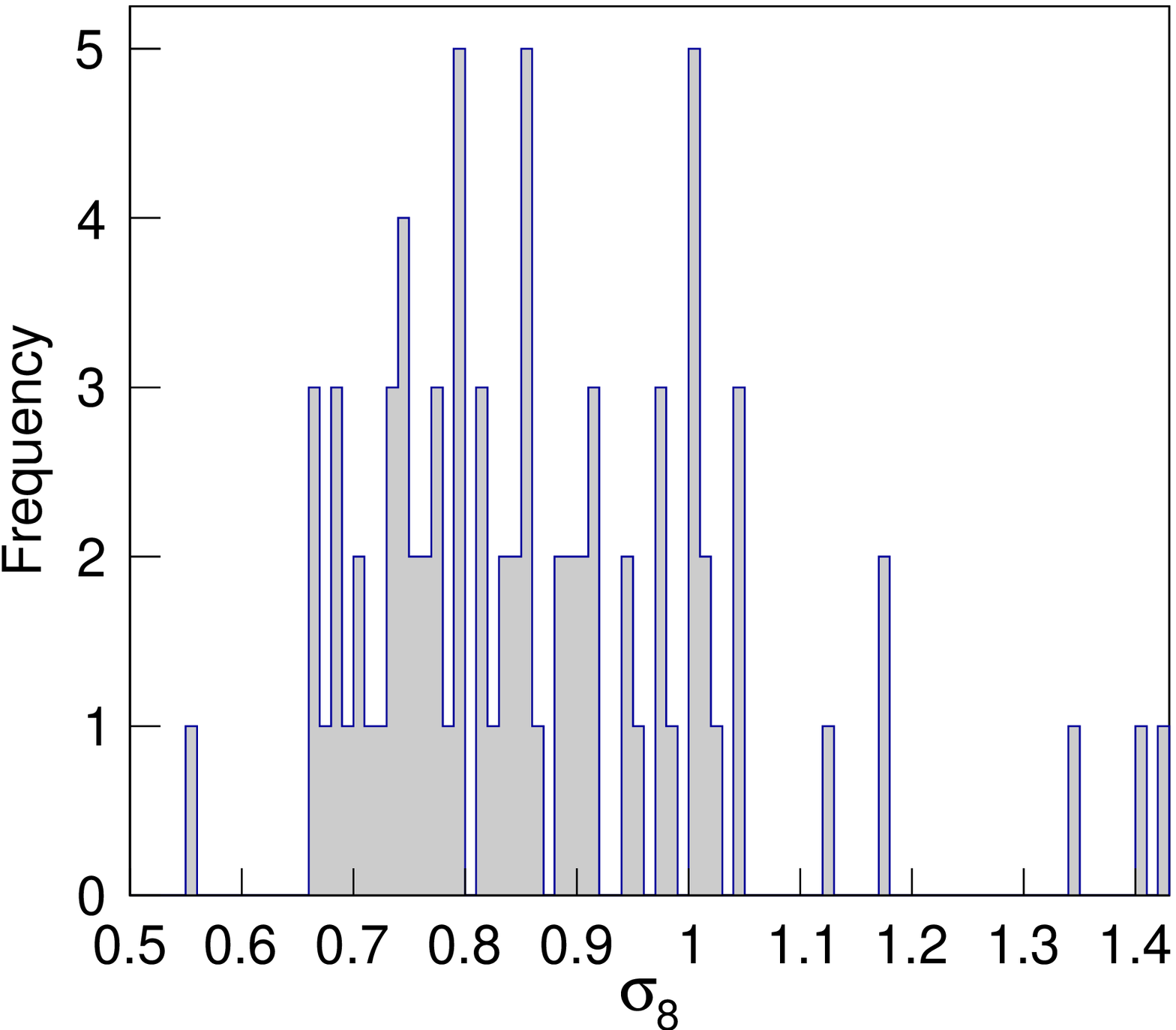}
\includegraphics[width=64mm,height=58mm]{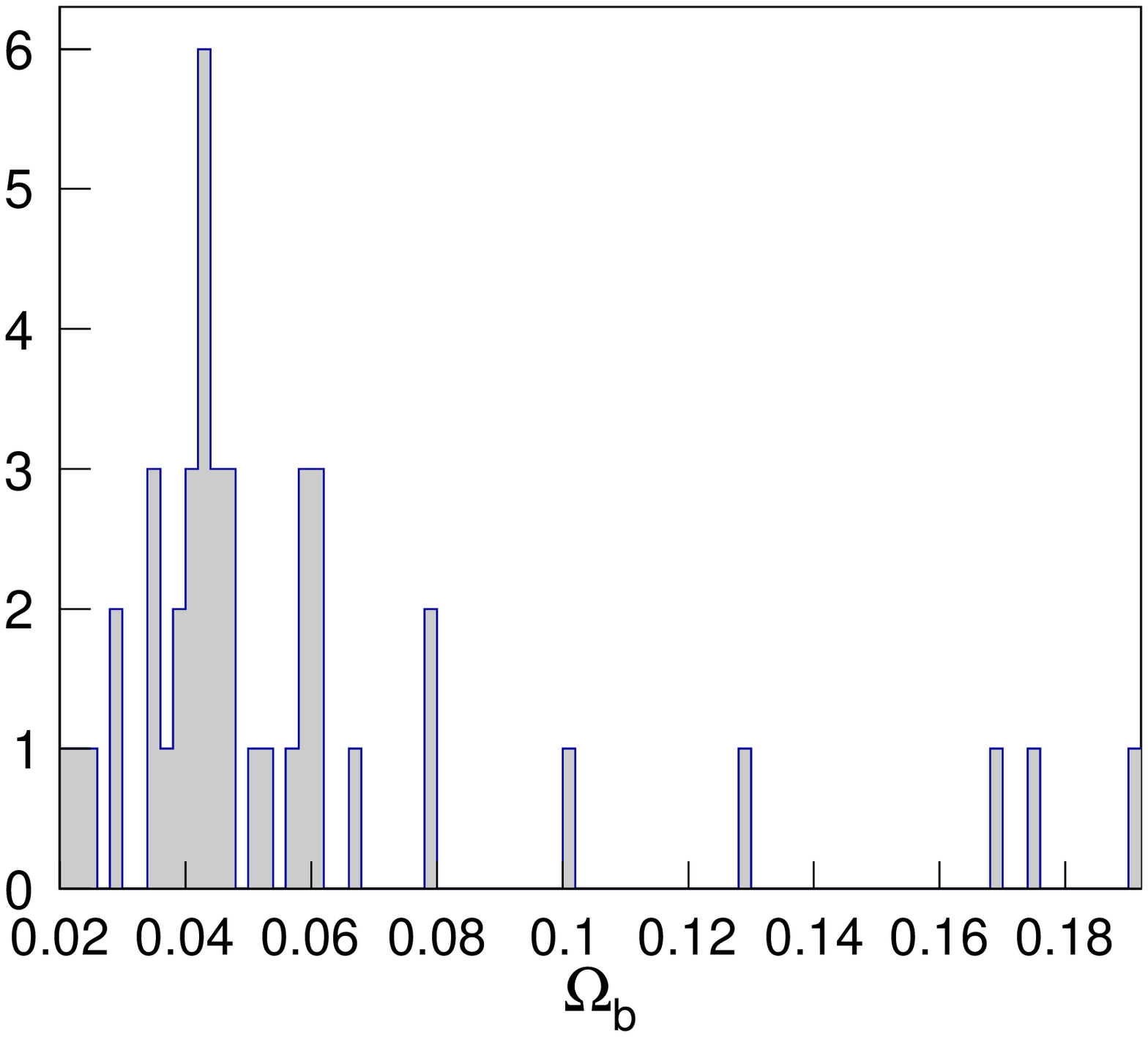}
\includegraphics[width=64mm,height=56mm]{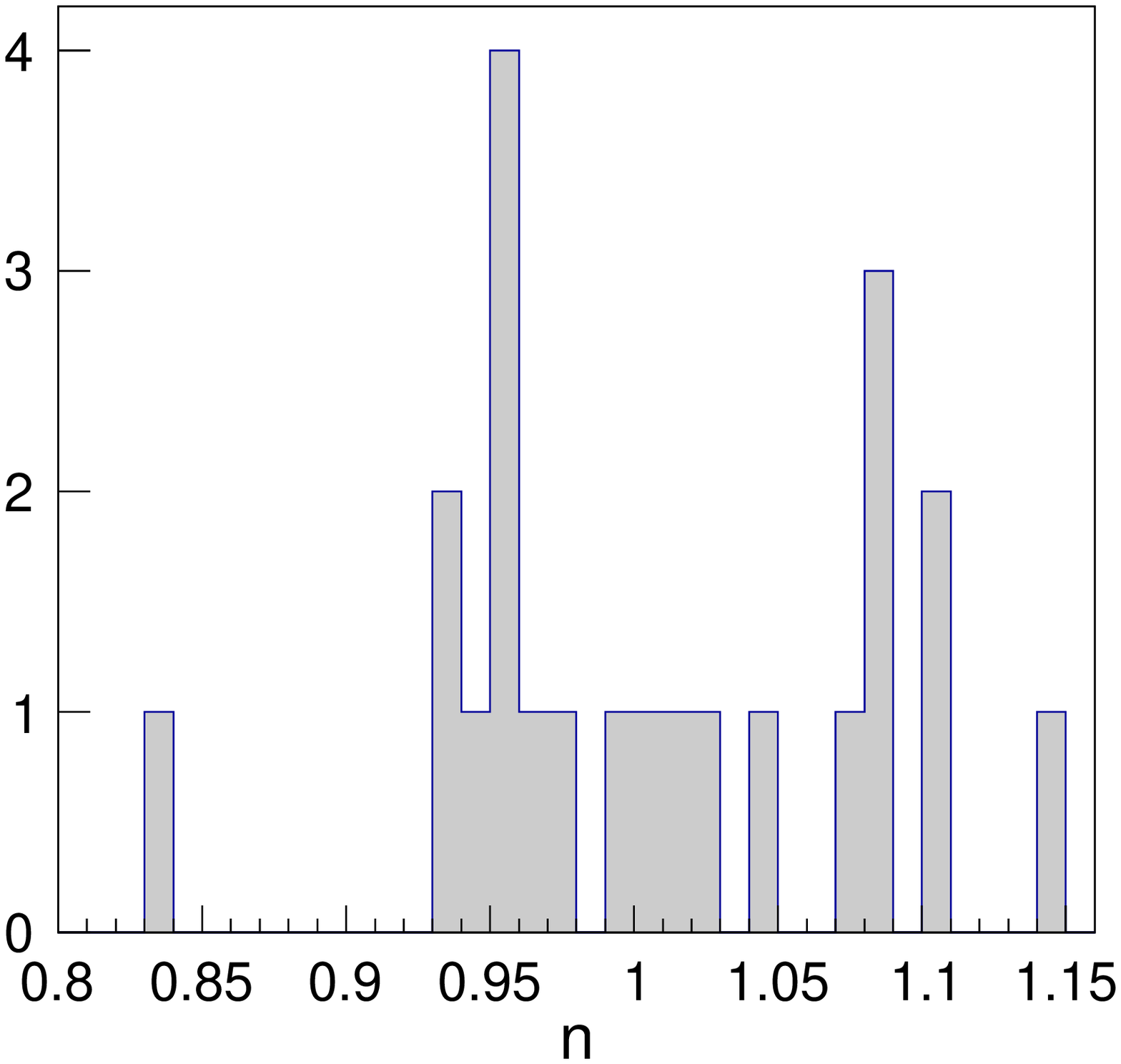}
\caption{Histograms of $\Omega_{m}$, $\Omega_{\Lambda}$, \& $h$ (top row, from left to right), and $\sigma_{8}$, $\Omega_{b}$, \& $n$ (bottom row, from left to right). Although used in our analyses, values of 39 for $\Omega_{m}$ and -1.5 for $n$ are not plotted. The bin size is 0.01 for all cases except for $\Omega_{b}$, where it is 0.001.} 
\label{Parameter 1-6 Histograms}
\end{figure}
\end{center}

\begin{center}
\begin{figure}[H]
\advance\leftskip-1.25cm
\advance\rightskip-1.25cm
\includegraphics[width=63mm,height=58mm]{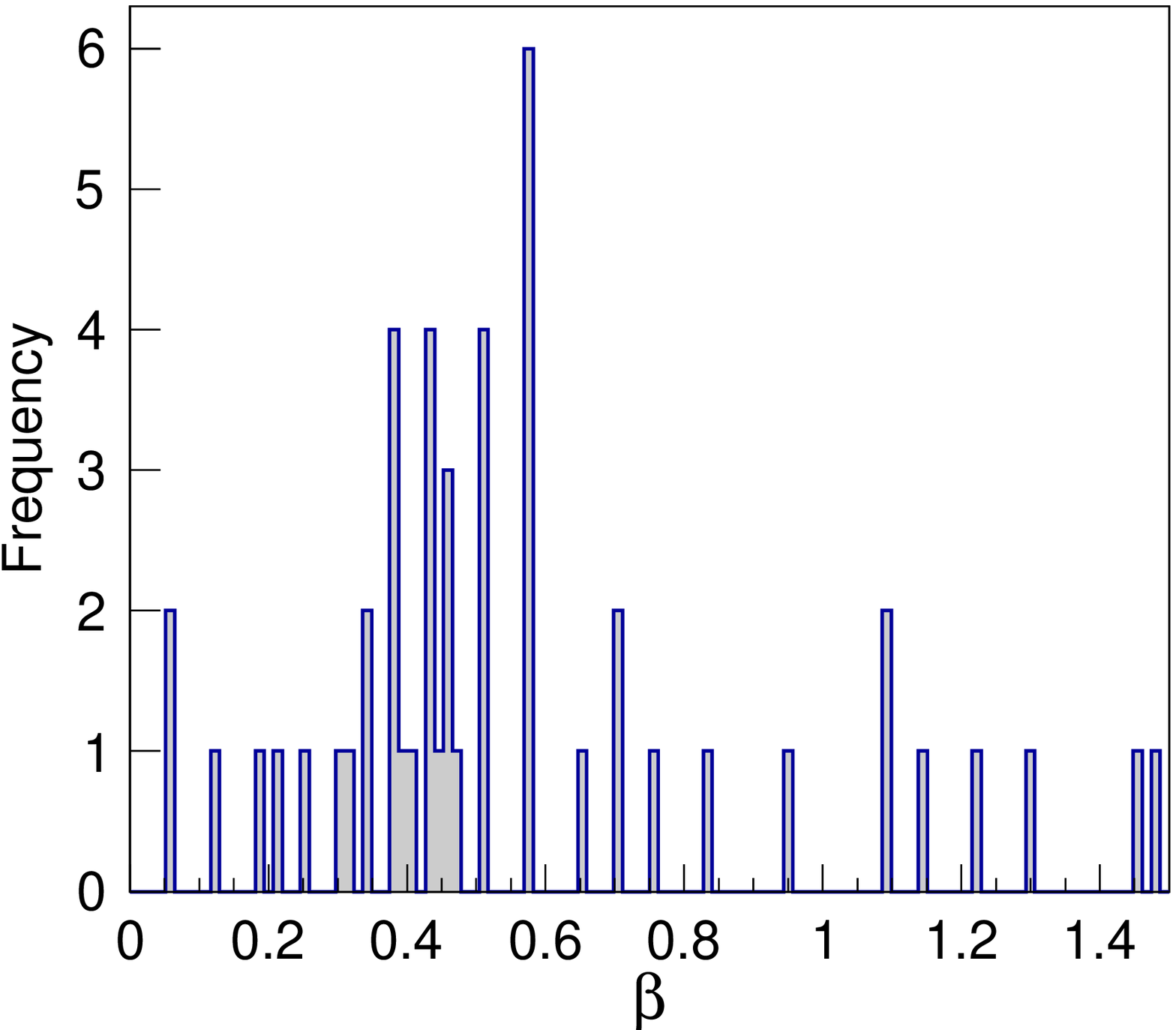}
\includegraphics[width=64mm,height=58mm]{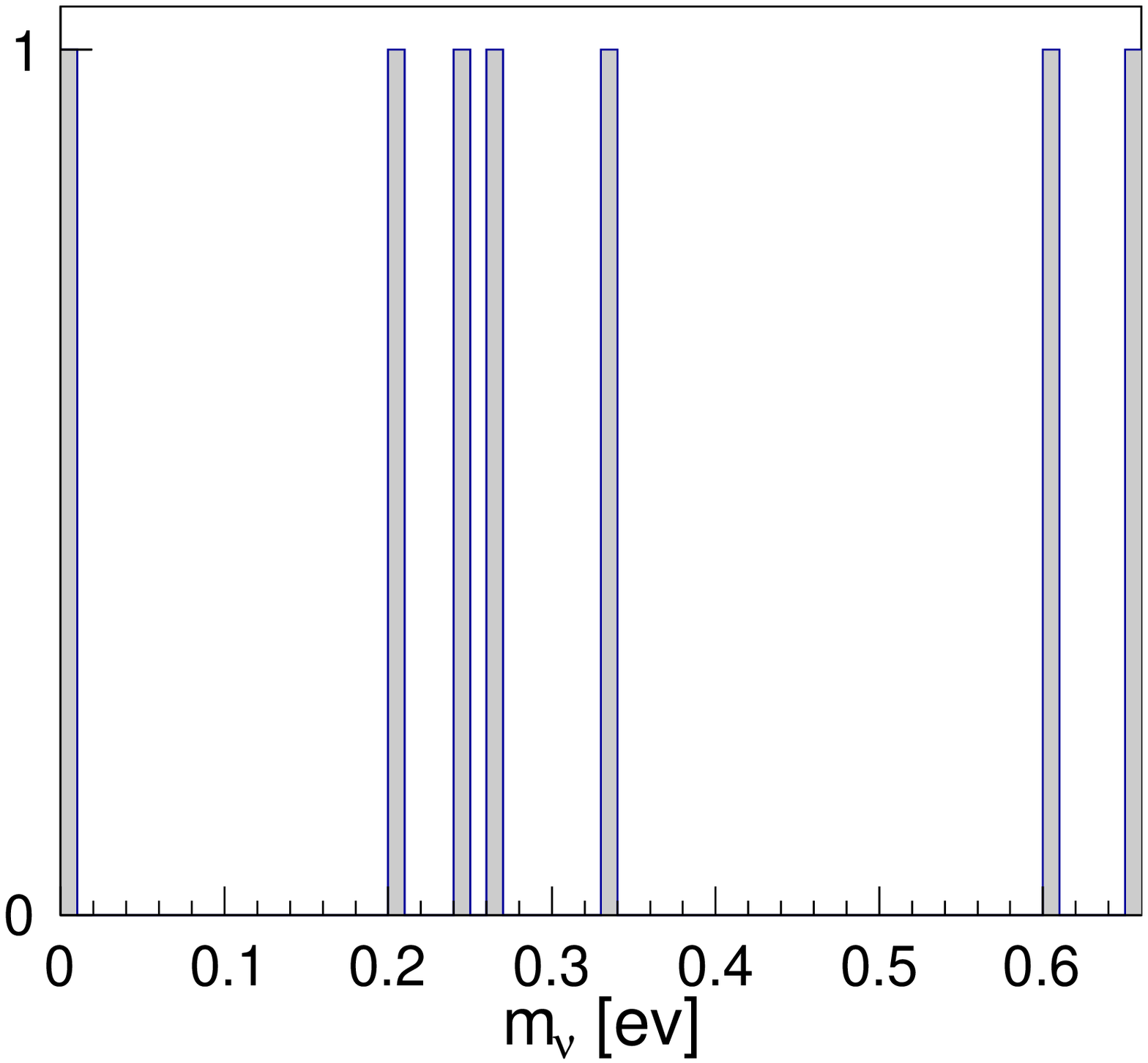}
\includegraphics[width=64mm,height=58mm]{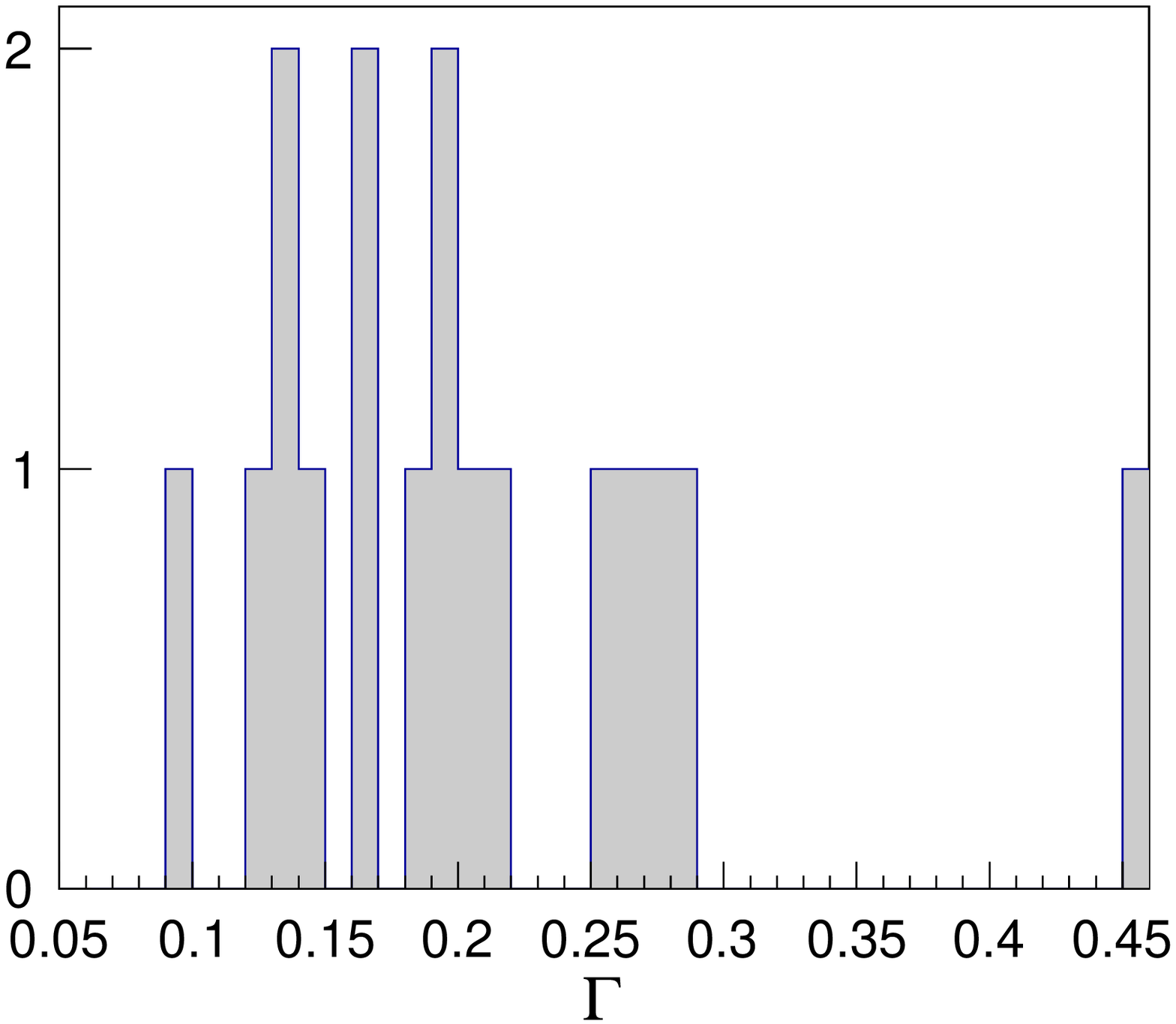}
\includegraphics[width=63mm,height=58mm]{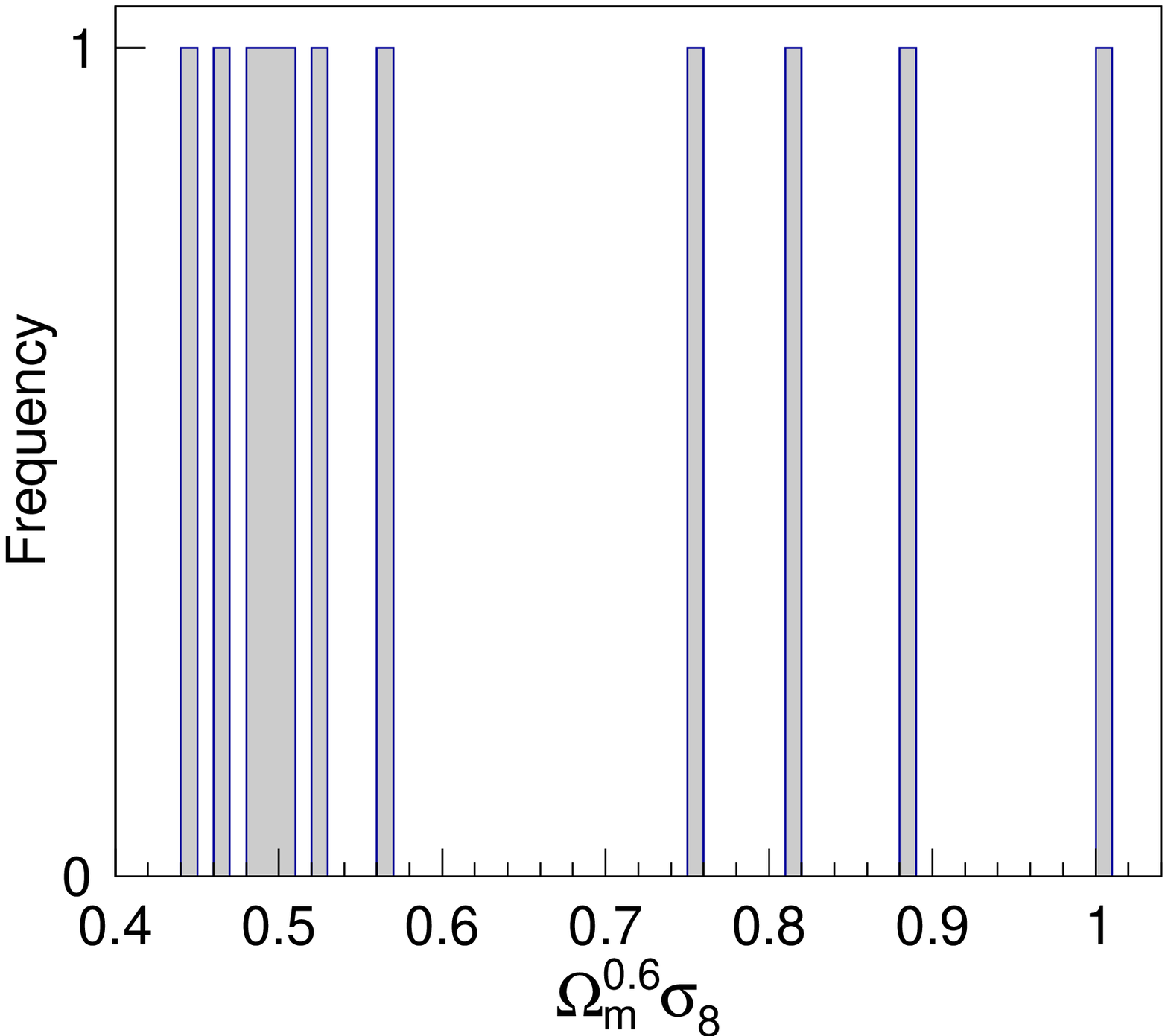}
\includegraphics[width=64mm,height=58mm]{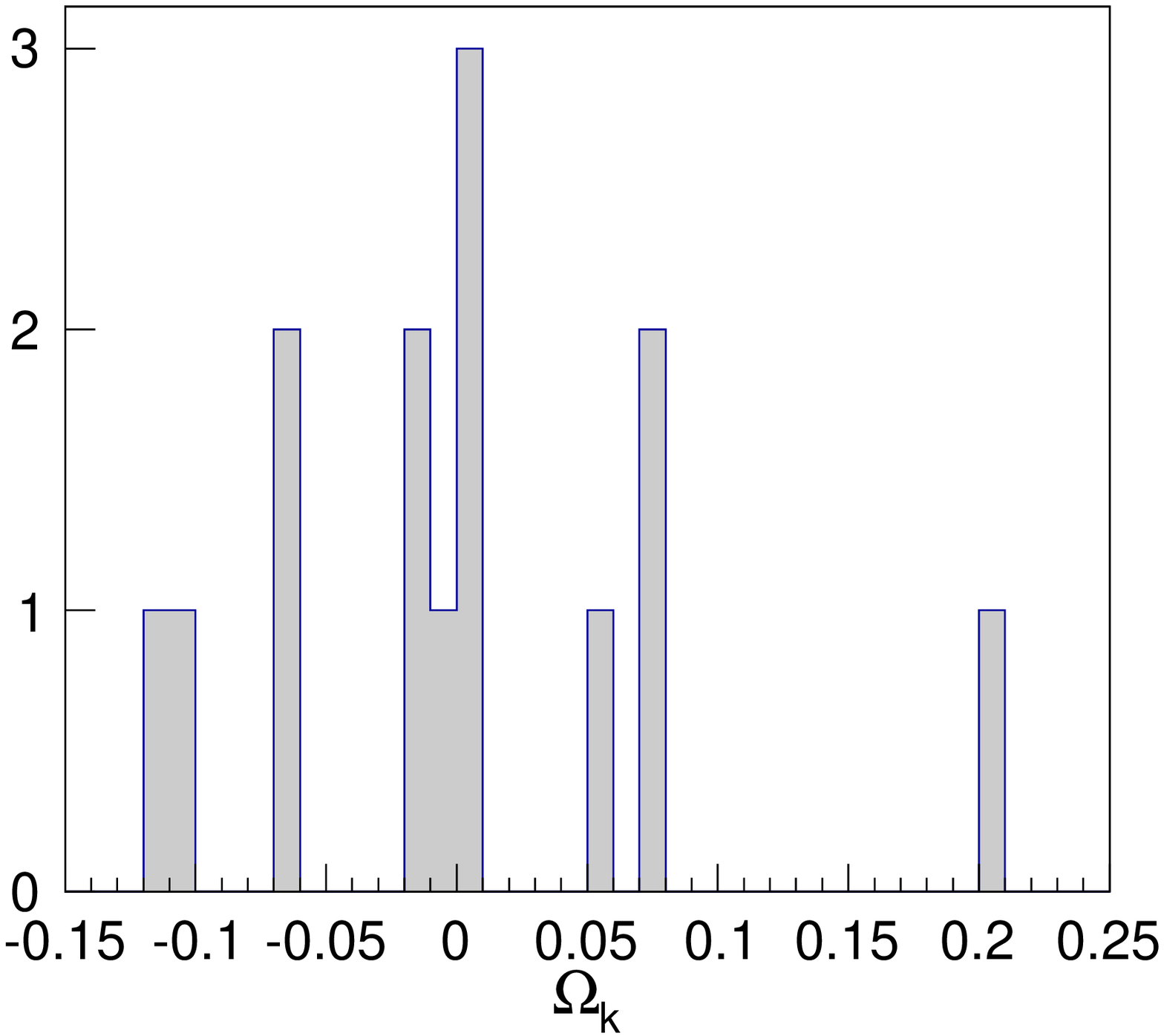}
\includegraphics[width=64mm,height=58mm]{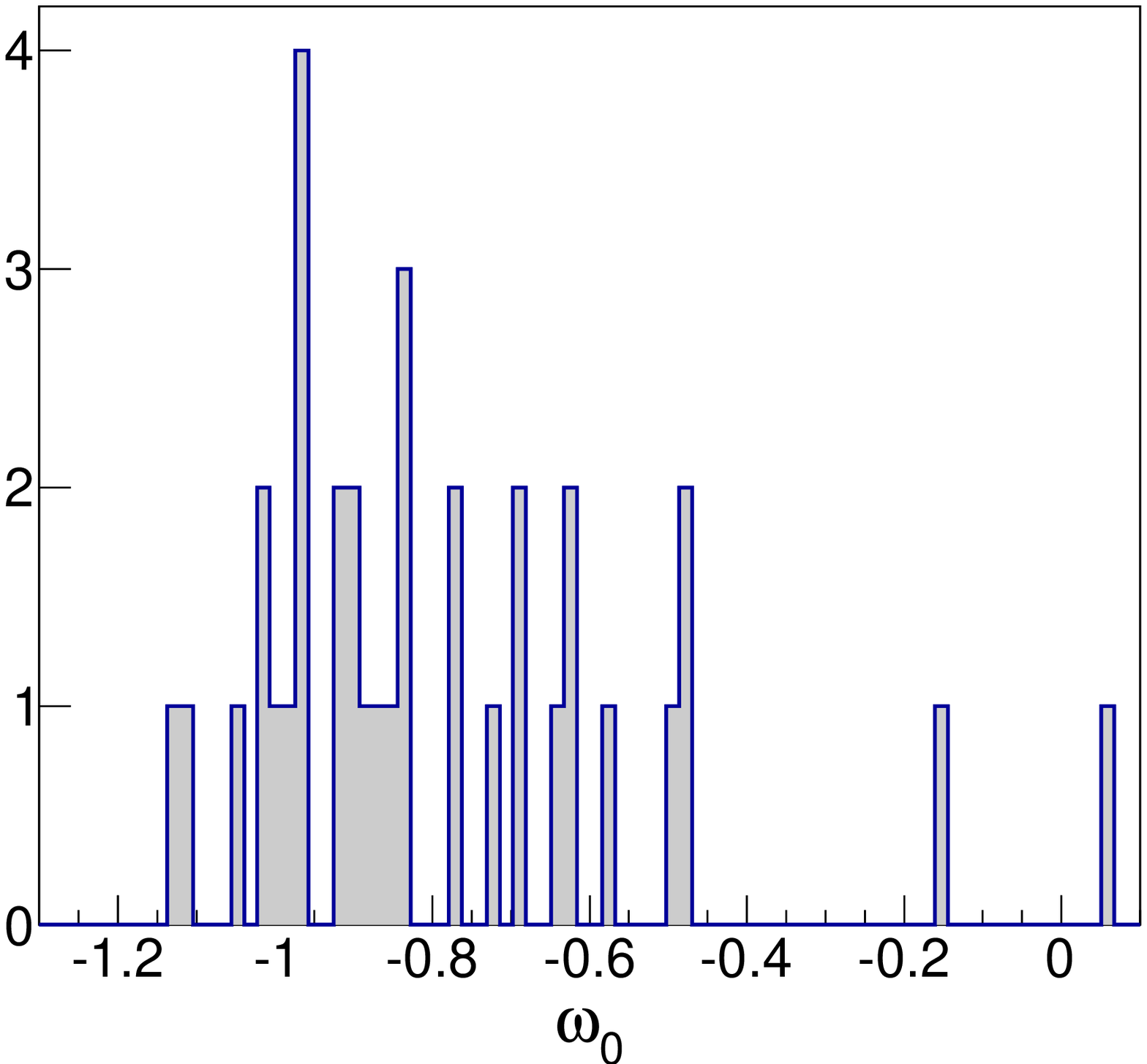}
\caption{Histograms of $\beta$, $m_{\nu}$, \& $\Gamma$ (top row, from left to right), and $\Omega_{m}^{0.6}\sigma_{8}$, $\Omega_{k}$, \& $\omega_{0}$ (bottom row, from left to right). Although used in our analyses, values of 2.48 ev for $m_{\nu}$ and 0.7 for $\Omega_{k}$ are not plotted. All of the above plots have a bin size of 0.01.} 
\label{Parameter 7-12 Histograms}
\end{figure}
\end{center}

Figures $\ref{Parameter 1-6 Histograms}$ and $\ref{Parameter 7-12 Histograms}$ show the 12 histograms of the 582 \cite{Croft2011} measurements. The histograms for parameters $\Omega_{k}$, $\Omega_{m}$, $m_{\nu}$, and $n$ have outlying values of 0.7, 39, 2.48 ev, and -1.5, respectively, omitted from their plots, though these values were used in our analyses. 

%%%%%%%%%%%%%%%
%figure 1
%%%%%%%%%%%%%%%

%%%%%%%%%%%%%%%
%figure 2
%%%%%%%%%%%%%%%

\section{Weighted Mean Statistics}
\label{WA Stat}
In analyzing data with known errors it is conventional to first consider a weighted mean statistic. This method yields a goodness of fit criterion that can be a valuable diagnostic tool. 

The standard formula  \citep[see, e.g.,][]{Podariu2001} for the weighted mean of cosmological parameter $q$ is
\begin{equation}
q_{\mathrm{wm}}=\frac{\sum_{i=1}^{N} q_{i}/\sigma_{i}^{2}}{\sum_{i=1}^{N}1/\sigma_{i}^{2}},
\end{equation}
where $q_{i}$ $\pm$ $\sigma_{i}$ are the central values and one standard deviation errors of the $i=1,2,...,N$ measurements. The weighted mean standard deviation of cosmological parameter $q$ is
\begin{equation}
\sigma_{\mathrm{wm}}=\left(\sum_{i=1}^{N} 1/\sigma_i^{2}\right)^{-1/2}. 
\end{equation}
One can also compute the goodness of fit $\chi^{2}$, 
\begin{equation}
\label{eq:chi}
\chi^2=\frac{1}{N-1}\sum_{i=1}^{N}\frac{(q_i-q_{\mathrm{wm}})^2}{\sigma_{i}^2}.
\end{equation}
Since this method assumes Gaussian errors, $\chi$ has expected value unity and error $1/\sqrt{2(N-1)}$. Hence, the number of standard deviations that $\chi$ deviates from unity is a measure of good-fit and is given as
\begin{equation}
\label{eq:N}
N_{\sigma}=|\chi-1|\sqrt{2(N-1)}.
\end{equation}
A large value of $N_{\sigma}$ could be an indication of unaccounted-for systematic error, the presence of correlations between the measurements, or the invalidity of the Gaussian assumption.

\section{Median Statistics}
\label{Med Stat}
The second statistical method we use is median statistics. This method makes fewer assumptions than the weighted mean method, and so can be used in cases when the weighted mean technique cannot. For a detailed description of the median statistics technique see \cite{Gott2001}\footnote{For recent applications of median statistics see, e.g., \cite{Shafieloo2011}, \cite{Barreira2011}, \cite{Rowlands 2011}, \cite{Pecaut2012}, \cite{Calabrese2012}, and \cite{Farooq2013a}.}. In summary, if we assume that the given measurements are: 1) statistically independent; and, 2) have no systematic error for the data set as a whole (as we also assume for weighted mean statistics), then as the number of measurements, $N$, increases to infinity, the median will reveal itself as a true value. This median is independent of measurement error \citep{Gott2001}, which is an advantage if the errors are suspect. This is also a disadvantage that results in a larger uncertainty for the median than for the weighted mean, because the information in the error bar is not used. 

If 1) is true then any value in the data set has a 50$\%$ chance of being above or below the true median value. As described in \cite{Gott2001}, if $N$ independent measurements $M_{i}$, where $i=1,...,N$, are taken then the probability of exactly $n$ measurements being higher (or lower) than the true median is 
\begin{equation}
P_{n}=\frac{2^{-N}N!}{n!(N-n)!}.
\end{equation}
It is interesting to note that for large $N$ the expectation value of the distribution width, $x$, of the true median is $\langle{x}\rangle=0.5$, with a standard deviation $\langle x^{2}-\langle{x}\rangle^{2}\rangle^{1/2}=1/(4N)^{1/2}$ \citep{Gott2001}. Of course, as $N$ increases to infinity, a Gaussian distribution is reached and median statistics recovers the usual standard deviation proportionality to $1/N^{1/2}$.

\section{Analysis}
\label{Analysis}
Since both weighted mean and median statistics techniques have individual benefits, we analyze the compilation of data for 12 parameters from \cite{Croft2011} using both methods. Our results are shown in Table \ref{table: WA and Med results}. Among other things, the table lists our computed weighted mean and corresponding standard deviation $\sigma_{\mathrm{wm}}$ value for the cosmological parameters, as well as the computed median value and the 1$\sigma$ and 2$\sigma$ intervals around the median.

Column 5 of Table \ref{table: WA and Med results} lists $N_{\sigma}$, the number of standard deviations the weighted mean goodness-of-fit parameter $\chi$ deviates from unity, see Eq.~(\ref{eq:N}). In all cases $N_{\sigma}$ is much greater than unity, indicating that the weighted mean results cannot be trusted. In the case of the Hubble constant this is likely due to the fact that the observed error distribution is non-Gaussian, see \cite{Chen2003a}.\footnote{The weighted mean technique also could not be used to combine different $\Omega_{m}$ measurements \citep{Chen2003b} or different cosmic microwave background temperature anisotropy observations \citep{Podariu2001}.} Perhaps a similar effect explains the large $N_{\sigma}$ values for some of the other parameters here. In any case, for our purpose here, the important point is that the weighted mean technique cannot be used to derive a summary estimate by combining together the different measurements tabulated by \cite{Croft2011} for each cosmological parameter.  

In a situation like this the median statistic technique can be used to combine together the measurements to derive an effective summary value of the cosmological quantity of interest \citep[e.g.,][]{Podariu2001, Chen2003b}. Column 6 of Table \ref{table: WA and Med results} lists the computed medians of the 12 cosmological parameters; the corresponding 1$\sigma$ and 2$\sigma$ ranges of these parameters are listed in columns 7 \& 8. 

The median statistics estimate for the Hubble parameter here, $h=0.68$ $_{-0.14}^{+0.08}$, is consistent with that estimated earlier by \cite{Chen2011} from 553 measurements of $h$ tabulated by Huchra, $h=0.68\pm0.028$ (with understandably much tighter error bars as a consequence of the many more measurements than the 124 we have used here).\footnote{For earlier, very consistent, estimates of $h$ using median statistics see \cite{Gott2001} and \cite{Chen2003a}.} Interestingly, from many fewer $\Omega_{m}$ measurements than considered here, \cite{Chen2003b} determine consistent, but somewhat tighter median statistics constraints on $\Omega_{m}$ by discarding the most discrepant, $\sim5\%$, of the measurements (those which contribute the most to $\chi^2$).

Also of interest, the median statistics estimates in Table \ref{table: WA and Med results} of $\Omega_{m}=0.29$ and $\sigma_{8}=0.84$ result in $\Omega_{m}^{0.6}\sigma_{8}=0.40$, which is significantly smaller than the median statistics estimate $\Omega_{m}^{0.6}\sigma_{8}=0.52$ listed in Table \ref{table: WA and Med results} that was determined directly from the 11 measurements of \cite{Croft2011}.\footnote{It is likely that the larger $\Omega_{m}^{0.6}\sigma_{8}=0.52$ found here is mostly a consequence of the higher  $\Omega_{m}^{0.6}\sigma_{8}$ values of a number of earlier analyses based on large-scale peculiar velocity measurements. While there are not enough measurements tabulated for us to more carefully examine this, it might be relevant that \cite{Croft2011} in the fifth paragraph of their Sec.~3.4, when discussing their Fig.~13, note that peculiar velocity measurements have not had a great track record when used to measure cosmological parameters.} On the other hand, $\Gamma=\Omega_{m}h$ computed using the median statistics estimates of $\Omega_{m}=0.29$ and $h=0.68$ is $\Gamma=0.20$, and is in very good agreement with the Table \ref{table: WA and Med results} median statistics value of $\Gamma=0.19$ from the 17 measurements of \cite{Croft2011}. 

In most cases the median statistics results of Table \ref{table: WA and Med results} provide reasonable (2010) summary estimates for the cosmological parameters. The one exception, perhaps, is that for $h$, which is estimated to be $h=0.68\pm0.028$ by \cite{Chen2011} from very many more measurements than the 124 used to derive the $h$ value in Table \ref{table: WA and Med results}. Perhaps the best current estimate of cosmological parameter values are those determined from the initial cosmic microwave background anisotropy measurements made by the $Planck$ satellite \citep{Ade2013}. The last two columns of Table \ref{table: WA and Med results} lists the $Planck$ estimates for most of these parameters. Here, the estimated cosmological constrained value and 1$\sigma$ standard deviation range (with the exception of $\Omega_{k}$ and $\omega_{0}$  that have 2$\sigma$ ranges, and $m_{\nu}$ that has a 2$\sigma$ upper limit) are listed.\footnote{The variance for parameters $\Gamma$ and $\Omega_{m}^{0.6}\sigma_{8}$ were not given in \cite{Ade2013}, but were calculated by adding their component's errors in quadrature (see the last footnote in Table \ref{table: WA and Med results}). All parameter estimates use both $Planck$ temperature power spectrum data as well as $WMAP$ polarization measurements at low multipoles. \cite{Ade2013} do not provide a $Planck$ estimate for $\beta$.}  
 
Comparing our computed median results to the recent $Planck$ values, one finds that almost all of the $Planck$ central value results fall within the 1$\sigma$ range of our median results. One exception is $\Omega_{m}^{0.6}\sigma_{8}$, possibly because of reasons discussed above; our estimates of $\Omega_{m}=0.29$ and $\sigma_{8}=0.84$ results in a $\Omega_{m}^{0.6}\sigma_{8}$ value which is very consistent with the $Planck$ estimate of $\Omega_{m}^{0.6}\sigma_{8}=0.415$. The other exception is $\omega_{0}$ which $Planck$ estimates to be -1.49. Our median statistics 2$\sigma$ range is $-1.25\leq\omega_{0}\leq-0.808$ computed from the 36 measurements of \cite{Croft2011}. \cite{Croft2011} note that the number of measurements for $\omega_{0}$ are still increasing with time\footnote{In fact, only around the time of $WMAP1$ were measurements, instead of limits, being published \citep{Croft2011}.}, unlike the case for the other parameters. As such, the estimation of $\omega_{0}$ is an area still under development and so we should not give much weight to the difference in our estimate from that of $Planck$.

More provocatively, it is instructive to compare our median statistics central estimates to the 1$\sigma$ (or 2$\sigma$) $Planck$ ranges. As expected, we see that our estimate of $\Omega_{m}$ ($\Omega_{\Lambda}$) lies somewhat above (below) the corresponding $Planck$ 1$\sigma$ range. Our estimates of $\Omega_{b}$ and $\Gamma$ are below the corresponding $Planck$ 1$\sigma$ ranges. Our estimate of $n$ is well above the $Planck$ 1$\sigma$ range, being quite consistent with the simplest scale-invariant spectrum \citep[]{Harrison1970, Peebles1970, Zeldovich1972} while $Planck$ data strongly favors a non-scale-invariant spectrum, also readily generated by quantum fluctuations during inflation \citep[see, e.g.,][]{Ratra1992}. And as might have been anticipated, our median statistics central $\Omega_{m}^{0.6}\sigma_{8}$ value is well above the $Planck$ 1$\sigma$ range.

%%%%%%%%%
%Table 1
%%%%%%%%%
\begin{sideways}
\begin{threeparttable}[H]
\caption{Weighted Mean and Median Statistics Results}
\vspace{4 mm}
\begin{tabular}{c c c c c c c c c c c c c} 
\hline\hline 
Parameter  & $N$\tnote{a} & WM\tnote{b} &  $\sigma_{\mathrm{wm}}$\tnote{c} & $N_{\sigma}$\tnote{d} & MS\tnote{e}  & 1$\sigma$ MS range\tnote{f} & 2$\sigma$ MS range\tnote{f} & ECV\tnote{g} & 1 or 2$\sigma$ range\tnote{h}\\
\hline
$\Omega_{m}$                              &  138    &   0.28     &  $3.8\times10^{-4}$  &  140   &  0.29   &  (0.21, 0.41)   &    (0.053, 0.76)  &  0.315  &  (0.297, 0.331)\\
$\Omega_{\Lambda}$                 &  38       &   0.72     &  $9.1\times10^{-4}$  &  30   &  0.72   &  (0.63, 0.77)   &    (0.47, 0.81)  &  0.685  &  (0.669, 0.703) \\
$h$                                                  &  124    &   0.63      &  $4.3\times10^{-4}$    &  160   &  0.68    &  (0.54, 0.76)   &   (0.41, 0.88)   &  0.673  &   (0.661, 0.685)\\
$\sigma_{8}$                                 &  80       &   0.86     &  $1.1\times10^{-3}$  &  130    &  0.84   &   (0.72, 1.0)  &   (0.56, 1.3)   &  0.829  &  (0.817, 0.841)\\
$\Omega_{b}$                               &  43       &   0.042   &  $1.8\times10^{-4}$  &    110    &  0.046   &   (0.031, 0.066)   &   (0.020, 0.17)   &  0.049  &  (0.048, 0.049)\\
$n$                                                   &  24      &   0.96      &  $9.2\times10^{-4}$  &  41  &  0.98   &  (0.94, 1.1)    &   (-1.5, 1.1)   &  0.960  &  (0.953, 0.968) \\
$\beta$                                            &  48       &   0.34     &  $2.9\times10^{-3}$  &  87   &  0.52   &   (0.39, 0.75)  &   (0.20, 1.2)   &\\
$m_{\nu}$[ev]                                &  8         &   0.014     &  $4.4\times10^{-3}$  &  16   &  0.26   &   (0.0070, 0.60)   &   (0.0, 0.65)  &   & \textless0.933\\
$\Gamma$                                      &  17       &   0.18     &  $4.1\times10^{-3}$  &  9.8   &  0.19   &   (0.13, 0.27)  &   (0.090, 0.45)  & 0.212  & (0.199, 0.223)\tnote{i}\\
$\Omega_{m}^{0.6}\sigma_{8}$ &  11       &   0.56    &  $1.1\times10^{-2}$  &  13    &  0.52   &  (0.46, 0.56)   &  (0.45, 0.57)   & 0.415	 &  (0.400, 0.427)\tnote{i}\\
$\Omega_{k}$                                &  15       &   $5.0\times10^{-3}$    &  $9.2\times10^{-4}$  &   23  &  0.0  &  (-0.091, 0.081)  &  (-1.1, 0.21)  & -0.037 &  (-0.086, 0.006)\\
$\omega_{0}$                                &  36       &  -0.968   &  $4.73\times10^{-4}$  &  51.9  &  -0.986   &  (-1.07, -0.808) &  (-1.25, -0.419)  & -1.49   &  (-2.06, -0.840)\\
\hline  
\end{tabular}
\begin{tablenotes}
\item[a]{Number of measurements.}
\item[b]{Weighted mean central value.}
\item[c]{Standard deviation of weighted mean.}
\item[d]{Number of standard deviations $\chi$ deviates from unity, Eq.~(\ref{eq:N})}
\item[e]{Median statistics central value.}
\item[f]{Median statistics range. In several cases for the 2$\sigma$ range there were not enough measurements to determine a 2$\sigma$ lower limit. In these cases, the lowest data point was used to represent the 2$\sigma$ lower limit. This is the case for $\Omega_{\Lambda}$, $\Omega_{b}$, $n$, $\beta$, $m_{\nu}$, $\Gamma$, $\Omega_{m}^{0.6}\sigma_{8}$, and $\Omega_{k}$.}
\item[g]{Estimated Constrained Value using $Planck$+WP ($WMAP$ polarization) data. These are from the last column of Table 2 of \cite{Ade2013}, except for $m_{\nu}$, $\Omega_{k}$, and $\omega_{0}$ which are from the third column of Table 10 in \cite{Ade2013}. For $m_{\nu}$ there was no central value listed and so a 2$\sigma$ upper limit is given. } 
\item[h]{Values are taken from Tables listed in the previous footnote. A 1$\sigma$ range was given for all parameters except for $m_{\nu}$, $\Omega_{k}$, and $\omega_{0}$ where a 2$\sigma$ upper limit or range is given.}
\item[i]{Here we have added in quadrature the errors on $\Omega_{m}$ and $h$ to get the range of $\Gamma$. To get the range for $\Omega_{m}^{0.6}\sigma_{8}$ we have taken the error on $\Omega_{m}^{0.6}$ which is given as $0.6\Omega_{m}^{0.4}\sigma_{\Omega_{m}}$ and added it in quadrature with the error on $\sigma_{8}$.} 
\end{tablenotes}
\label{table: WA and Med results}
\end{threeparttable}
\end{sideways}

\section{Conclusion}
\label{Conclusion}
From the measurements compiled by \cite{Croft2011}, the median statistics technique can be used to compute summary estimates of 12 cosmological parameters. On comparing 11 of these values to those recently estimated by the $Planck$ collaboration, we find good consistency in 9 cases. The two exceptions are the parameters $\Omega_{m}^{0.6}\sigma_{8}$ and $\omega_{0}$. It is likely that the $Planck$ estimate of $\Omega_{m}^{0.6}\sigma_{8}$ is more accurate, while $\omega_{0}$ estimation is still in its infancy and so one should not give much significance to this current discrepancy. 

It is very reassuring that summary estimates for a majority of cosmological parameters considered by \cite{Croft2011} are very consistent with corresponding values estimated from the almost completely independent $Planck$ + $WMAP$ polarization data. This provides strong support for the idea that we are now converging on a ``standard'' cosmological model.

\acknowledgments

We are grateful to Rupert Croft for giving us the \cite{Croft2011} data and for useful advice. We also thank Omer Farooq for helpful discussions and useful advice. This work was supported in part by DOE grant DEFG03-99EP41093 
and NSF grant AST-1109275.

\end{document}